\documentclass{article}
\usepackage{spconf,amsmath,graphicx}
\usepackage{booktabs}
\usepackage{marvosym}
\usepackage{fancyhdr}

\title{ACT-Net: Asymmetric Co-Teacher Network for Semi-supervised Memory-efficient Medical Image Segmentation}
\name{Ziyuan Zhao$^{\dag\ddag\sharp}$, Andong Zhu$^{\S}$, Zeng Zeng$^{\dag,\ddag}$\textsuperscript{,\Letter}, 
Bharadwaj Veeravalli$^{\S}$, Cuntai Guan$^{\sharp}$}

\address{
$^{\dag}$Institute of Infocomm Research (I$^2$R), A*STAR, Singapore\\
$^{\ddag}$ Artificial Intelligence, Analytics And Informatics (AI$^3$), A*STAR, Singapore\\
$^{\sharp}$School of Computer Science and Engineering, Nanyang Technological University, Singapore \\
$^{\S}$School of Electrical and Computer Engineering, National University of Singapore, Singapore 
}
\begin{document}
%\ninept
%
\maketitle

\thispagestyle{fancy}
\fancyhead{}
\lhead{}
\vspace{-0.5pt}
\lfoot{\footnotesize{Copyright 2022 IEEE. Published in 2022 IEEE International Conference on Image Processing (ICIP), scheduled for 16-19 October 2022 in Bordeaux, France. Personal use of this material is permitted. However, permission to reprint/republish this material for advertising or promotional purposes or for creating new collective works for resale or redistribution to servers or lists, or to reuse any copyrighted component of this work in other works, must be obtained from the IEEE. Contact: Manager, Copyrights and Permissions / IEEE Service Center / 445 Hoes Lane / P.O. Box 1331 / Piscataway, NJ 08855-1331, USA. Telephone: + Intl. 908-562-3966.}}
\cfoot{}
\rfoot{}
\begin{abstract}

While deep models have shown promising performance in medical image segmentation, they heavily rely on a large amount of well-annotated data, which is difficult to access, especially in clinical practice. On the other hand,  high-accuracy deep models usually come in large model sizes, limiting their employment in real scenarios. In this work, we propose a novel asymmetric co-teacher framework, ACT-Net, to alleviate the burden on both expensive annotations and computational costs for semi-supervised knowledge distillation. We advance teacher-student learning with a co-teacher network to facilitate asymmetric knowledge distillation from large models to small ones by alternating student and teacher roles, obtaining tiny but accurate models for clinical employment. To verify the effectiveness of our ACT-Net, we employ the ACDC dataset for cardiac substructure segmentation in our experiments. Extensive experimental results demonstrate that ACT-Net outperforms other knowledge distillation methods and achieves lossless segmentation performance with $250\times$ fewer parameters.


\end{abstract}

%%%%%%%%%%%%%%%%%%%%%%%%%%%%%%%%%%%%%%%%%%%%%%%%%%%%%%%%%%%%%%%%%%%%%%%%%%%%%%%%
\section{INTRODUCTION}

Recently, deep learning~(DL) has become increasingly popular in the field of medical image segmentation~\cite{long2015fully,ronneberger2019u}. With large-scale, well-annotated datasets, deep learning models can perform reliably in a variety of medical imaging tasks~\cite{zhao2019bira,LIU2020244}. However, labeling such datasets is time-consuming and labor-intensive, making it impractical in some domains, particularly in real-world clinical scenarios. In this regard, various label-efficient DL methods~\cite{tajbakhsh2020embracing}, including semi-supervised learning~\cite{tepens2017,meanteacher}, self-supervised learning~\cite{zeng,hu2021semi}, and active learning~\cite{yang2017suggestive,zhao2021dsal} have been developed, achieving great success with limited labels on medical image segmentation.

On the other hand, most DL models are computationally expensive and memory intensive, impeding the deployment of large DL models on resource-constrained devices or applications with strict latency requirements,~\emph{e.g.}, real-time mobile health (mHealth) applications. Therefore, various methods for model compression have been proposed to lower the computational requirements, including parameter pruning, quantization~\cite{zhang2021medq,hajabdollahi2018low}, and knowledge distillation~\cite{hinton2015distilling}. Among these, model distillation~\cite{hinton2015distilling} has recently made significant progress in terms of transferring knowledge between small student and big teacher models in order to improve the accuracy of the small model. However, in low annotation regimes, suboptimal big models can negatively impact the performance of small models, failing to satisfy clinical employment. These motivate us to develop a deep learning framework to address both label scarcity and high computational expense problems.

In this work, we propose a unique asymmetric co-teacher network, called ACT-Net, for label- and memory-efficient learning in an end-to-end manner. In ACT-Net, a large teacher model and a small student model are generated in a teacher-student learning manner for knowledge distillation. To further address the label scarcity, we build on recent progress in mean-teacher training~\cite{meanteacher} by constructing a self-ensembling co-teacher network using unlabeled data to train more accurate big and small student models, thereby facilitating better knowledge distillation. Experimental results on the ACDC dataset validate the effectiveness of the proposed ACT-Net on low-label model compression regimes, achieving comparable segmentation performance to big models with much fewer parameters. The main contributions can be summarized:
\vspace{0pt}
\begin{itemize}
  \vspace{-4pt}\item To address both label scarcity and model complexity, a new asymmetric co-teaching architecture called ACT-Net is proposed. It contains a teacher-student network for model distillation and a co-teacher network for semi-supervised learning.
  \vspace{-4pt}\item We design an asymmetric co-teaching strategy to facilitate both heterogeneous and homogeneous knowledge distillation from teacher and co-teacher models.
  \vspace{-4pt}\item Extensive experiments and analysis were conducted, demonstrating the advances of the proposed methods for model compression under label scarcity.
\end{itemize}

\begin{figure*}[!thb]
\centering
\includegraphics[width=0.75\textwidth]{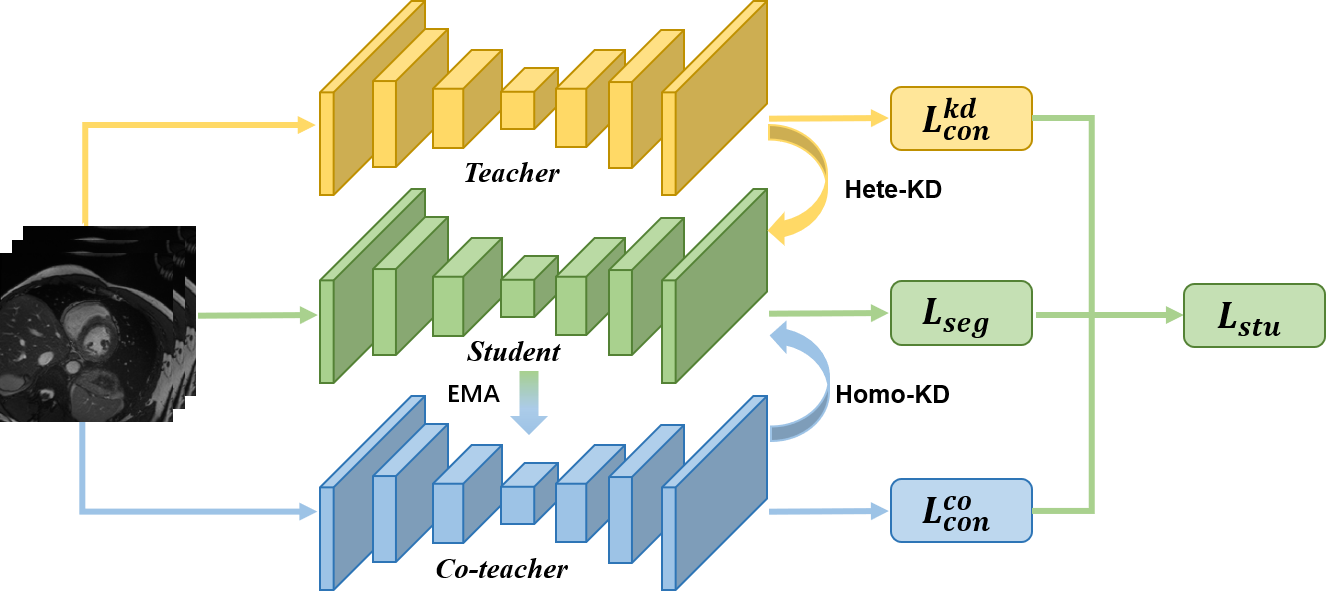}

%  \caption{The overall structure of our framework:(a) The red line, the green line and the blue line show the data flow with annotated image, unannotated image, and ground truth, respectively. (b) $\tau$=1 means the output is processed by softmax and $\tau$ \textgreater 1 represents that the output is processed by modified softmax of knowledge distillation with specific $\tau$. In our experiment $\tau$ = 3 (c). Only annotated images contribute to the supervised loss.  For unannotated images, transformation consistency and knowledge distillation consistency form unsupervised loss.}
 \caption{The overall architecture of ACT-Net. The student model learns from both labeled data and unlabeled data via Heterogeneous Knowledge Distillation~(Hete-KD) and Homogeneous Knowledge Distillation~(Homo-KD) simultaneously for label-efficient and memory-efficient segmentation} %At the same time EMA is applied between student and co-teacher model.}
\label{fig:pipeline}
\end{figure*}
\section{RELATED WORK}
%
% Ronneberger~\emph{et al.} proposed U-Net~\cite{ronneberger2019u} in 2015, whose segmentation performance is better than the previous FNC~\cite{long2015fully} suggested by Long~\emph{et al.}.Subsequent work based on the U-Net structure~\cite{LIU2020244}, such as the extension of the attention module to the network to get Attention-U-Net~\cite{oktay2018attention} and ensemble U-Net~\cite{kamnitsas2017ensembles}, has demonstrated the general applicability and good performance of U-Net in the field of medical image segmentation. 
% Therefore,modified U-Net network is selected as our base model to implement image segmentation task.

To alleviate the reliance on annotations, plenty of deep learning methods other than strictly supervised learning have been proposed for label-efficient medical image segmentation~\cite{tajbakhsh2020embracing,zeng,yang2017suggestive, media2022urpc}, among which, semi-supervised learning based on self-ensembling~\cite{tepens2017,meanteacher} has recently attracted much attention due to its powerful capacity on leveraging unlabeled data, achieving promising segmentation results. More specifically, Laine and Aila~\cite{tepens2017} first introduced Temporal Ensembling to encourage prediction consistency across different training steps. After that, the mean teacher~(MT) framework~\cite{meanteacher} was proposed based on the exponential moving average (EMA) weights of the student model to force the student-teacher output consistency. Following the spirit of teacher-student learning, many works further extended MT with different constraints for medical image segmentation tasks~\cite{yu2019uncertainty,li2021hierarchical,luo2021semi} in the label-scarce scenarios. However, the computational complexity of deep learning introduces another vital issue for medical applications in clinical practices.

Knowledge distillation (KD) is a frequently used technique for model compression in resource-constrained circumstances, in which, a compact student model is trained with the help of a big teacher model for the same data.
% For model compression in the resource-constrained scenarios, knowledge distillation~(KD) is one of the most widely applicable techniques, in which, a compact student model is trained with the help of a big teacher model for the same data. 
Hinton~\emph{et al.}~\cite{hinton2015distilling} first proposed the concept of knowledge distillation, introducing hard and soft label constraints for model training. Following that, many works have been established for model compression from different perspectives, including response-based distillation~\cite{chen2017learning,cho2019efficacy}, feature-based distillation~\cite{kim2018paraphrasing,heo2019comprehensive}, and relation-based distillation~\cite{peng2019correlation}. However, these methods require large-scale labeled datasets, restricting their applications in medical image segmentation with limited annotations. It is noted that both self-ensembling methods and KD strategies leverage teacher-student training for their respective domains, whereas the former concentrates on knowledge distillation from self-ensembling models with the same architecture, and the latter transfers knowledge between heterogeneous networks. Inspired by these observations, we propose an asymmetric co-teacher network for semi-supervised memory-efficient medical image segmentation.

\section{METHODOLOGY}

Given annotated data $x^s$ and unannotated data $x^u$, Our goal is to leverage both $x^s$ and $x^u$ to improve semi-supervised performance and compress the model for deployment. The overview of the proposed ACT-Net is illustrated in Fig.~\ref{fig:pipeline}, consisting of a teacher model, a co-teacher model, and a student model. For model compression, the small student model is taught by the big high-accuracy pretrained teacher model via knowledge distillation. To ease the effects of label scarcity, a co-teacher network with the weights updated from the student model via EMA is constructed to uncover the hidden knowledge beneath unannotated data. The teacher and the co-teacher jointly transfer the knowledge to the student model for label-efficient memory-efficient segmentation.

% The architecture is shown in Figure~\ref{fig:pipeline}. $X^S$ and $X^U$ are annotated and unannotated image respectively. $N$ represents random noise. 

% Three models are prepared for our framework, two pre-trained models, a teacher model $\theta^T$ with high accuracy and a co-teacher model $\theta^{CO}$ with the same structure of student model $\theta^S$. The teacher model $\theta^T$ is pre-trained via the mean teacher method. Then transfer knowledge from teacher model to co-teacher model $\theta^{CO}$ through knowledge distillation. Co-teacher model generated by knowledge distillation outperforms models with the same structure, trained by the mean teacher method~\cite{meanteacher}. The detailed training process is mentioned in the subsequent methodology section.

% In order to learn a compact and fast DNN model that can
% be adopted in mobile applications, we propose a stagewise knowledge distillation method. This method enables
% the small student model to capture not only the information
% in ground truth labels but also the information distilled from
% the cumbersome teacher model.

\subsection{Heterogeneous Knowledge Distillation}
To transfer knowledge across different architectures, we follow the generalized knowledge distillation strategy~\cite{hinton2015distilling} to generate well-informed soft labels from the teacher to guide the student model. First, same inputs $x_i$ are fed into the student model $f_{s}$ and teacher model $f_{t}$ to generate soft predictions:
\begin{equation}
P_{i}^s=\sigma\left(f_{s}\left(x_i; \theta_s\right) / \tau\right), 
P_{i}^t=\sigma\left(f_{t}\left(x_i; \theta_t\right) / \tau\right), \nonumber
\end{equation}
where $\sigma$ is the softmax function, $\theta_s$ and $\theta_{t}$ are the weights of the student and teacher, respectively, and $\tau$ is the temperature parameter used for smoothing the predicted probabilities, which is set to $20$, empirically. To force the student to imitate the teacher behavior for better performance, we encourage the predictions of the student model $f_s\left(\theta_s\right)$ and the teacher $f_t\left(\theta_t\right)$  model to be consistent by minimizing the knowledge distillation loss $L_{con}^{kd}$ defined as:
\begin{equation}
\label{eq:loss_kd}
L_{con}^{kd}=\sum_{i=1}^{N}\left\|P_{i}^s-P_{i}^t\right\|_{2},
\end{equation}
where $N$ is the number of images and we apply the mean squared error (MSE) loss for regularizing the consistency.

\subsection{Homogeneous Knowledge Distillation}
Despite the knowledge distillation from teacher to student, both teacher and student models suffer from performance degradation in the label-scarce scenarios. To address label scarcity, we design the co-teacher model $f_{c}$ following the self-ensembling strategy~\cite{{tepens2017,meanteacher}} to leverage unlabeled data for boosting model performance. To be more specific, the co-teacher model for the student model share the same architecture with the student $f_{s}$, and its weights $\theta_c$ are updated by EMA of $\theta_{s}$ in different training epochs $t$:
\begin{equation}
\theta_{c}^t=\alpha \theta_{c}^{t-1}+(1-\alpha) \theta_{s}^t,\nonumber
\end{equation}
where $\alpha$ governs the effects of the current parameter in the student model. We feed the inputs with different perturbations to the student and the co-teacher, respectively, and we expect the predictions to be consistent. In this regard, the consistency regularization is introduced by reducing the difference with the MSE loss denoted as:
\begin{equation}
\label{eq:loss_co}
L_{con}^{co}=L_{con}^{co}(f_s\left(x; \theta_s, \xi \right), f_c\left(x; \theta_c, \xi^{\prime}\right)),
\end{equation}
where $\xi $ and $\xi^{\prime}$ denote the different perturbations to the inputs of student and co-teacher, respectively. Similarly, we can also train the big teacher model by means of self-ensembling to obtain a high-accuracy pretrained teacher model for model compression. For simplicity, we omit the self-ensembling process of the teacher model in Fig.~\ref{fig:pipeline}.

\subsection{Asymmetric Co-teaching Strategy}
The student model in our ACT-Net learns from labeled data $x^x$ using the supervised loss $L_{seg}$, defined as:
\begin{equation}
\label{eq:loss_seg}
L_{seg}= L_{dice}(f_s\left(x^s; \theta_s\right), y^s) + L_{ce}(f_s\left(x^s; \theta_s \right), y^s),
\end{equation}
where $L_{dice}$ and $L_{ce}$  respectively signify dice loss and cross-entropy loss. The hybrid loss function is commonly utilized in the segmentation task for medical images~\cite{hu2021semi,li2021hierarchical}. In addition, the student model concurrently distill the knowledge from the teacher model and the co-teacher model for both semi-supervised learning and model compression. Finally, by integrating losses from Eq.~(\ref{eq:loss_kd}), Eq.~(\ref{eq:loss_co}), and Eq.~(\ref{eq:loss_seg}), for the student model, the intact loss function can be formatted as:
\begin{equation}
\begin{array}{r}
L_{stu}=\underbrace{L_{seg}}_{\text {supervised}} + \underbrace{\lambda_{con}^{kd}L_{con}^{kd}}_{\text {heterogeneous}} + \underbrace{\lambda_{con}^{co} L_{con}^{co}}_{\text {homogeneous}},\nonumber
\end{array}
\end{equation}
where $\lambda_{con}^{kd}$ and $\lambda_{con}^{co}$ are the trade-off hyperparameters of $L_{con}^{kd}$ and $L_{con}^{co}$, respectively. With the proposed asymmetric co-teaching strategy, the student model can explore both annotated data and unannotated data in an end-to-end manner for label-efficient memory-efficient segmentation.

\section{EXPERIMENTS}

\subsection{Dataset and Experimental Settings}

We extensively evaluated the proposed framework on ACDC dataset~\cite{bernard2018deep} for cardiac segmentation. The public training dataset with $100$ cases was used in our experiments and randomly split into training set ($70$), validation set ($10$), and testing set ($20$). $10\%$ (7 cases) of training set are used as labeled data. Three cardiac substructures were included for segmentation,~\emph{i.e.}, left ventricle (LV), right ventricle (RV), and myocardium (MYO). For pre-processing, slices were extracted from the transverse plane, resized to $256\times256$, and then normalized to bring all values into the range [0,1]. The dice similarity coefficient~(DSC)~\cite{li2021hierarchical} was applied for performance comparison, and a higher DSC means better results.

U-Net~\cite{ronneberger2019u}, a flexible U-shape encoder-decoder architecture, was selected as our backbone, allowing us to easily adjust the network structure and parameter numbers for our model compression experiments. Concretely, we denote the number of encoder layers as $L$, and the first layer in encoder has $N_1$ output channels, and the depth of feature map in the $i$-th layer can be defined as $N_i =N_1\times 2^{\left(i-1\right)}$. In this regard, $L$ and $N_1$ can be used to adjust the depth and the width of the U-Net,~\emph{e.g.,} a $5$-layer U-Net with $32$ initial channels can be represented by $\text{U-Net}_{[5,32]}$. SGD optimizer is set to update with momentum $0.9$. The initial learning rate was set to $0.01$ with the learning rate schedule $lr^t=lr \times \left(1-t/t_{max}\right)^{0.9}$, where $t$ and $t_{max}$ are the current number of iterations and the maximum iterations, respectively. The EMA decay rate $\alpha$ was $0.99$. The batch size was set $20$, consisting of $10$ labeled data and $10$ unlabeled data. Both $\lambda_{con}^{kd}$ and $\lambda_{con}^{co}$ were set to $0.5$, empirically. In our experiments, we chose the big model $\text{U-Net}_{[6,64]}$ and the small model $\text{U-Net}_{[4,16]}$ for model compression. We first pretrained the big and small models with self-ensembling for $30$k iterations, separately. Then, the whole architecture was trained for $30$k iterations. 

\subsection{Results and Discussions}

\begin{table}[!thb]
\centering
\setlength\tabcolsep{3pt}
\caption{Segmentation results of different methods.}
\scalebox{0.8}{
\begin{tabular}{c|c|c|c|c|c|c} 
\toprule
\begin{tabular}[c]{@{}c@{}}\\Method\end{tabular} & Network       & \% Labels & RV              & MYO             & LV              & Mean             \\ 
\hline
FS-upper                                         & $\text{U-Net}_{[6,64]}$ & 100         & 0.9138          & 0.8862          & 0.9401          & 0.9133           \\
FS-upper                                         & $\text{U-Net}_{[4,16]}$ & 100         & 0.9090          & 0.8843          & 0.9358          & 0.9093           \\
FS-lower                                         & $\text{U-Net}_{[6,64]}$ & 10          & 0.8043          & 0.8125          & 0.8598          & 0.8255           \\
FS-lower                                         & $\text{U-Net}_{[4,16]}$ & 10          & 0.7892          & 0.7973          & 0.8555          & 0.8140           \\ 
\hline
MT                                               & $\text{U-Net}_{[6,64]}$ & 10          & 0.8298          & 0.8438          & 0.8898          & 0.8544           \\
MT                                               & $\text{U-Net}_{[4,16]}$ & 10          & 0.8094          & 0.7995          & 0.8648          & 0.8297           \\ 
\hline
KD                                               & $\text{U-Net}_{[4,16]}$ & 100         & 0.9143          & 0.8857          & 0.9409          & 0.9137           \\
KD                                               & $\text{U-Net}_{[4,16]}$ & 10          & 0.8292          & 0.8171          & 0.8764          & 0.8391           \\ 
\hline
\textbf{ACT-Net}                                 & $\text{U-Net}_{[4,16]}$ & 10          & \textbf{0.8505} & \textbf{0.8284} & \textbf{0.8880} & \textbf{0.8556}  \\
\bottomrule
\end{tabular}}
\label{mytable_model}
\end{table}

\begin{table}[!thb]
\centering

\caption{Complexity comparison of different models}
\label{size}
\scalebox{0.75}{
\begin{tabular}{c|c|c|c} 
\hline
\ Structure  & Params~(M) &Model size~(MB)  &FLOPs~(G)\\ 

\hline
U-Net$_{[4,16]}$        &\textbf{0.45 }& \textbf{1.74 } &  4.6                                           \\
% $U-Net_[5,16] $             & $1.81$ & 6.95                                                          \\ 
U-Net$_{[5,32]}$              & 7.24 & 27.69  &   23.56                                                    \\ 
% $U-Net_[5,64]$              & 28.96 & 110.53                                                       \\ 
U-Net$_{[6,64]}$             & \textbf{116.01} & \textbf{442.66   }   &115.56       \\ 

\hline
\end{tabular}}
\label{tab:complexity}
\end{table}

\subsubsection{Comparison with different methods}
\begin{figure}[t]
\centering
\includegraphics[width=0.42\textwidth]{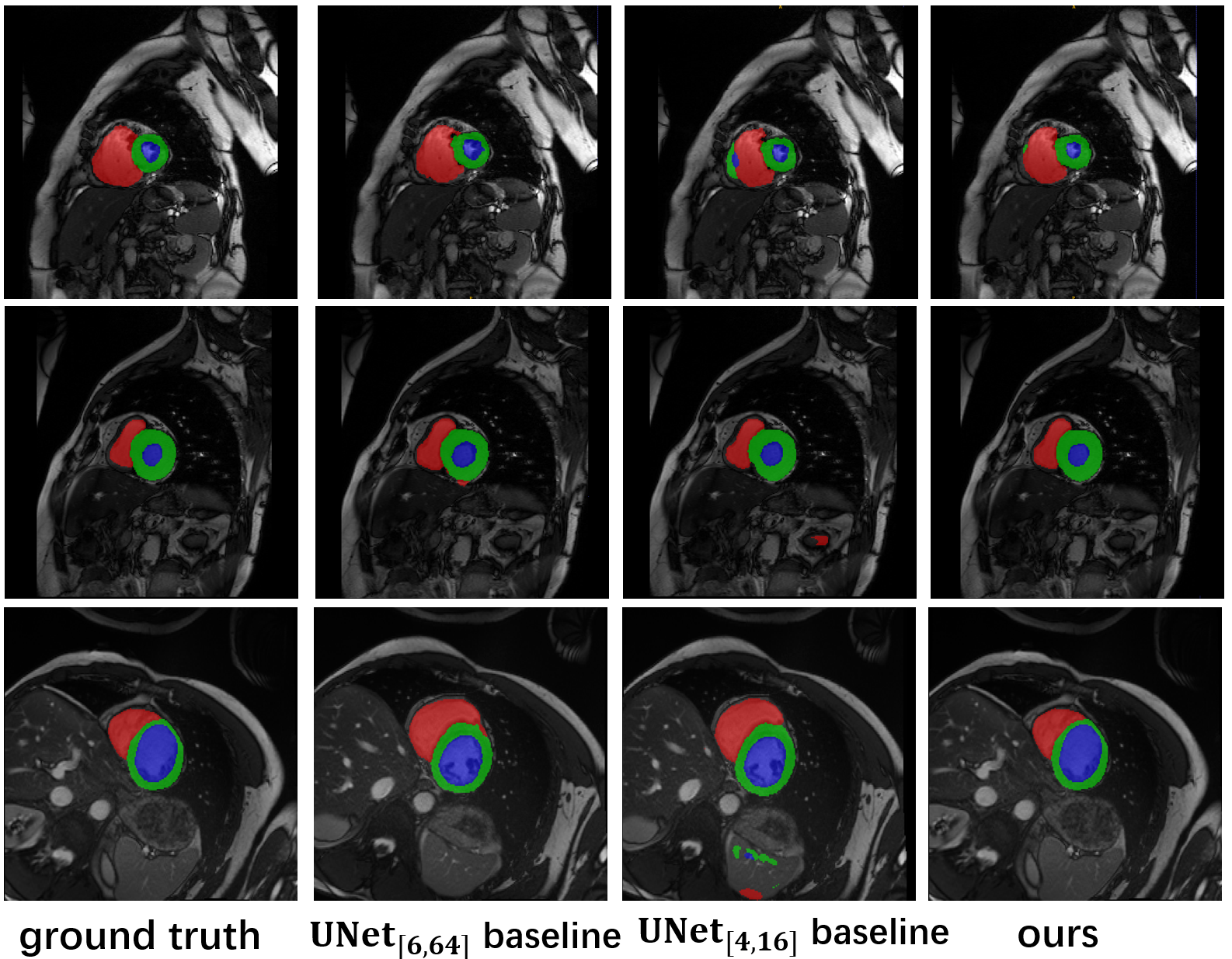}

 \caption{Visualization of results by different methods.}
\label{fig:visual}
\end{figure}

We trained the model with only labeled data for baselines, referred to as fully supervised~(FS), and also compared our framework with different methods, including mean-teacher~(MT)~\cite{meanteacher} and knowledge distillation~(KD)~\cite{chen2017learning}. We used different architectures and ratios of labels for a comprehensive comparison. The comparisons between different methods are shown in Table~\ref{mytable_model}. The big model tends to have better performance than the small model; however, the performance of both models drops seriously when using only $10\%$ annotations. The mean-teacher~(MT) framework can leverage unlabeled data to boost performance with different architectures, and the big model has more considerable improvements than the small model, which suggests the high-capacity model can also benefit semi-supervised learning. From table~\ref{tab:complexity}, it is observed that the big model is nearly $250\times$ larger than the small one. For model compression, knowledge distillation~(KD) can achieve comparable performance with the small model using fully annotated data; however, when using only $10\%$ annotations, the performance of the small model is significantly decreased. In comparison, our method can generate more precise predictions with the small model and $10\%$ annotations, even outperforming the MT (U-Net$_{[6,64]}$, $10\%$). The visualization results in Fig.~\ref{fig:visual} reveal that our method can provide more accurate segmentation results with fewer errors in segmentation.

% \begin{table}[thb]
% \centering
% \label{{tab:complexity}}
% \caption{Complexity comparison of different models}
% \label{size}
% \scalebox{0.8}{
% \begin{tabular}{c|c|c|c} 
% \hline
% \ Structure  & Params~(M) &Model size~(MB)  &FLOPs~(G)\\ 

% \hline
% U-Net$_{[4,16]}$        &\textbf{0.45 }& \textbf{1.74 } &  4.6                                           \\
% % $U-Net_[5,16] $             & $1.81$ & 6.95                                                          \\ 
% U-Net$_{[5,32]}$              & 7.24 & 27.69  &   23.56                                                    \\ 
% % $U-Net_[5,64]$              & 28.96 & 110.53                                                       \\ 
% U-Net$_{[6,64]}$             & \textbf{116.01} & \textbf{442.66   }   &115.56       \\ 

% \hline
% \end{tabular}}
% \end{table}

\subsubsection{Ablation analysis}
\begin{figure}[!thb]
\centering
\includegraphics[width=0.47\textwidth]{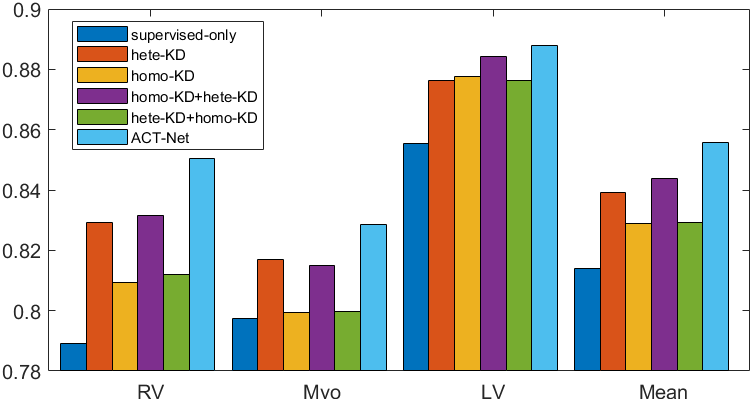}

 \caption{Ablation results of different components.}
\label{fig:bar}
\end{figure}

The ablation results of different components of the proposed ACT-Net (U-Net$_{[4,16]}$, $10\%$) are illustrated in Fig.~\ref{fig:bar}. Both heterogeneous knowledge distillation~(hete-KD) and homogeneous knowledge distillation~(homo-KD) can help improve the performance, especially when there are few labels. However, a simple sequential combination of these two KD methods,~\emph{e.g.,} hete-KD + homo-KD, cannot lead to better results, which may cause negative transfer for the student model. Moreover, it is noted that the performance of homo-KD + hete-KD is similar to hete-KD. We hypothesize that the reason for this is that the knowledge distillation from teacher to student (hete-KD) is heavily limited by label scarcity, despite the well-trained teacher using homo-KD.
In contrast, ACT-Net can effectively integrate different knowledge distillations in the same training stage to boost segmentation performance. By combining the advantages of both homo-KD and hete-KD, ACT-Net (U-Net$_{[4,16]}$, $10\%$) can achieve $85.56\%$ in mean dice, outperforming other methods on semi-supervised memory-efficient segmentation.

\section{CONCLUSIONS}
In this work, we propose an asymmetric co-teacher network to solve label scarcity and model complexity. We promote both heterogeneous and homogeneous knowledge distillation from various teachers in order to achieve label- and memory-efficient medical image segmentation. Extensive experiments and analysis conducted on the ACDC dataset demonstrate the effectiveness of the proposed method on the label-scarce resource-constrained scenarios, and the proposed method is easily adaptable to other segmentation tasks for clinical deployments.

\clearpage
\bibliographystyle{IEEEbib}
\bibliography{refs1.bib}

\end{document}